\RequirePackage{amsmath}
\documentclass[runningheads]{llncs}
\usepackage[T1]{fontenc}
\usepackage{graphicx}
\usepackage{cite}
\usepackage{algorithm}
\usepackage{amssymb}
\usepackage{algpseudocode}
\usepackage{hyperref}
\usepackage{caption}
\usepackage{subcaption}
\usepackage{url}
\usepackage{breakurl}
\usepackage{color}

\begin{document}
\title{The Liouville Generator for Producing Integrable Expressions}
%
%
\author{Rashid Barket\inst{1}\orcidID{0000-0002-9104-4281} \and 
Matthew England\inst{1}\orcidID{0000-0001-5729-3420} \and
J\"{u}rgen Gerhard\inst{2}}
\authorrunning{R. Barket et al.}
%
\institute{Coventry University, Coventry, United Kingdom\\
\email{\{barketr, matthew.england\}@coventry.ac.uk} 
\and 
Maplesoft, Waterloo, Ontario, Canada\\
\email{jgerhard@maplesoft.com}
}
%
\maketitle              
\begin{abstract}
There has been a growing need to devise processes that can create comprehensive datasets in the world of Computer Algebra, both for accurate benchmarking and for new intersections with machine learning technology. We present here a method to generate integrands that are guaranteed to be integrable, dubbed the LIOUVILLE method. It is based on Liouville's theorem and the Parallel Risch Algorithm for symbolic integration. 
We show that this data generation method retains the best qualities of previous data generation methods, while overcoming some of the issues built into that prior work. The LIOUVILLE generator is able to generate sufficiently complex and realistic integrands, and could be used for benchmarking or machine learning training tasks related to symbolic integration.  

\end{abstract}

\section{Introduction}

The problem of how to generate expressions that are known to be integrable has become important for various applications. For one, to support the emerging interactions between the fields of Machine Learning (ML) and symbolic integration: these range from performing algorithm selection within computer algebra systems \cite{Sharma2023_SIRD, Barket2024TreeLSTM} to having an ML model predict an integral directly \cite{Lample2020}. Going a step further, there are now general-purpose ML models built to undertake various mathematics tasks at a university level which include definite and indefinite integration \cite{Drori2022_mathNet, Hendrycks2021_MATHdata}. Most ML models require training on an extensive amount of data, so being able to generate many integrable expressions would be a benefit to all these ML projects.

Computer Algebra Systems (CASs) are also seeing improvements year after year without making use of ML. Maple, a prominent CAS, has seen several new methods made available to users for indefinite integration in recent years\footnote{\url{https://www.maplesoft.com/products/maple/new_features/maple2023/index.aspx}} for example. Elsewhere we have seen a new CAS, wxMaxima, developed for calculus teaching which includes integration in its system \cite{Karjanto2021_wxmaxima}. Similarly, Irvanian et al. recently added integration functionality to the symbolic manipulation packages of SciML for Julia \cite{Iravanian2022_sciML}; and the Rule-Based Integrator (RUBI) \cite{Rich2018_RUBI} has now been implemented as the main integrator for the CAS SymJa (Symbolic Java), and also has an implementation in SymPy (a CAS for Python). All of these, and many other CASs that are being implemented/receiving updates, would benefit from having a comprehensive set of integrands available for thorough benchmarking of their (in)definite integration commands. 

Only a handful of data generation methods currently exist for creating (integrand, integral) pairs. Lample \& Charton \cite{Lample2020} created three methods to produce these pairs which each have their own shortcomings as we will soon see in Section~\ref{sec:existing}. The present authors recently presented an alternative data generator based upon the Risch Algorithm \cite{Barket2023_generation}. While that method does address some of the issues in the first three methods from \cite{Lample2020}, it is limited in scope to what types of integrands it can produce. The drawbacks of all of these methods necessitate the need for another data generator to ensure a rich variety of indefinite integrable expressions. This led to our new data generator, which we dub LIOUVILLE.

This paper focuses on how LIOUVILLE can generate integrable expressions that other methods have difficulty producing.  The paper continues in Section \ref{sec:existing} with an overview of the existing methods.  Then in Section \ref{sec:Background} we give the background material necessary to present the new method in Section \ref{sec:New}.  We analyse the new method and clarify its benefits in Section \ref{sec:discussion}, before concluding in Section \ref{sec:Conc}.

\section{Existing Data Generation Methods}\label{sec:existing}

The five known methods for producing integrable expressions are discussed here, along with their shortcomings.

\subsection{Deep Learning for Symbolic Mathematics}

Lample \& Charton created three data generation methods in \cite{Lample2020} (whose title is that of this subsection) for their application of ML to predict the integrals of given expressions:
\begin{itemize}
    \item \textbf{FWD:} Integrate an expression $f$ through a CAS to get $F$ and add the pair ($f, F$) to the dataset.
    \item \textbf{BWD:} Differentiate $f$ to get $f'$ and add the pair ($f', f$) to the dataset.
    \item \textbf{IBP:} Given two expressions $f$ and $g$, calculate $f'$ and $g'$. If $\int f'g$ is known (i.e. exists in a database) then the following holds (integration-by-parts):
    $\int fg' = fg - \int f'g$.
    Thus, we add the pair ($fg'$, $fg - \int f'g$) to the dataset. 
\end{itemize}

These offer a starting point but do not generate the rich variety of data that we need. Discussions of the shortcomings of these methods have already been published by the present authors \cite{Barket2023_generation}, and others \cite{Piotrowski2019_critique, Davis2019_critique}. We give a high-level overview here. 

First, for both the FWD and IBP methods we find they end up generating pairs of short integrands and long integrals or vice versa \cite{Barket2023_generation} while pairs of balanced length are possible (and arguably more common) in practice.  

Second, for all three data generation methods, we find they also end up producing many \emph{similar} integrands. By \emph{similar}, we mean expressions that are the same up to the coefficients in the expression \cite{Piotrowski2019_critique}.  This means that a high number of examples does not necessarily mean a high range of the space of interest is sampled, and it also runs the risk of data leakage between the training and testing sets of any ML experiment upon the dataset. 

Third, these datasets seem to be missing integrands that look hard, but in reality, are easily solvable. Consider the following integral from \cite{Davis2019_critique}:
\begin{equation}\label{eq:easy_int}
    \int \sin^2(e^{e^{x}}) + \cos^2(e^{e^{x}}) dx
\end{equation}
Integral (\ref{eq:easy_int}) looks complicated but the integrand is just equivalent to the number $1$ so the integral evaluates simply to $x$. Almost any CAS would be able to ascertain this but the ML model would have a hard time as the data generation methods are extremely unlikely to generate data of this sort. Theoretically, the FWD method could, but it is statistically unlikely.

Lastly, we highlight certain types of integrands that the BWD method is extremely unlikely to generate. Consider the expression
\begin{equation}\label{eq:setup}
    f_1 = \frac{1}{\log(x)} + \frac{1}{\log^2(x)}.
\end{equation}
If we were to generate data using BWD, we would then differentiate (\ref{eq:setup}) to get 
\begin{equation}\label{eq:diff_setup}
    f_{1}' = -\frac{1}{\log^{2}(x) x}-\frac{2}{\log^{3}(x) x}.
\end{equation}
Notice that in this partial fraction representation of the answer in (\ref{eq:diff_setup}), none of the fractions have a denominator of degree 1 in $\log(x)$. If we desire this, we need to add a logarithm to Equation (\ref{eq:setup}) in the following way.

\begin{flalign}
    &f_2 = f_1 + \log(\log(x)) \label{eq:setup_new}\\
    &f_{2}' = -\frac{1}{\log^{2}(x) x}-\frac{2}{\log^{3}(x) x} + \frac{1}{x\log(x)} \label{eq:diff_setup_new}
\end{flalign}

The addition of $\log(\log(x))$ in Equation (\ref{eq:setup_new}) provided the derivative in Equation (\ref{eq:diff_setup_new}) with a linear degree logarithm denominator in the partial fraction representation of the answer. It is unlikely that any random expression generator will produce the additional logarithms needed to get a partial fraction answer where one of the fractions has degree 1. The BWD method will almost always create integrals that have a degree 2 or higher denominator whenever the integrand includes a denominator.       

\subsection{Generating Elementary Integrable Expressions}\label{sec:risch_gen}

The present authors created the RISCH data generation method in \cite{Barket2023_generation} (whose title was the same as this subsection) to help address some of the issues that Lample \& Charton faced with their data generation methods. RISCH was based on the Risch algorithm \cite{Risch1969_original} and Liouville's theorem (presented in Section \ref{sec:liouville}).

RISCH addresses two issues the FWD, BWD, and IBP methods are not able to overcome. First, the bias that BWD and FWD suffer on the lengths of the expressions (either short or long integrands and the opposite for the integral). The RISCH method can partially address this issue because of the structure of Liouville's theorem: we can access the parameters available when generating an integrand to address the bias. Some parameters must be constants, but if the starting denominator is not square-free, we also get parameters that can be functions. For such parameters, we can choose the functions to be as big or as small as desired. 


Our work in \cite{Barket2023_generation} also sought to address the second shortcoming set out above, namely avoiding similar expressions within the dataset. Recall that two expressions are considered similar if they only differ by their coefficients. Again, by making use of those functional parameters in the RISCH generation method we can prevent this: we simply choose the functional parameters to be sufficiently different and the issue of similar expressions is avoided.  It is possible that more care over the random generation parts of the methods in \cite{Lample2020} could also have avoided this issue.

However, the RISCH generator does introduce different types of limitations not seen in the first three generators. 
\begin{enumerate}
    \item The Risch algorithm allows elementary extensions that can be either logarithmic, exponential, or algebraic. The first two cases are relatively straightforward and are what we implemented for the RISCH generator in \cite{Barket2023_generation}. In theory, one could make the RISCH generator work for algebraic extensions but would require a lot more development time. This is also true of adapting it to include special functions.
    \item In the partial fraction representation, only denominators of maximum degree 2 were included in order to avoid complications arising from non-linear polynomial solving in higher degrees.
    \item The algorithm in \cite{Barket2023_generation} works by considering a tower of extensions and solving each sub-case recursively. Similarly to point 1., more extensions are possible in theory but are complex to implement in practice. Our implementation of the RISCH generator does not support towers of extension.  
\end{enumerate}   

\subsection{The Substitution Rule}

The substitution rule from calculus is a technique used to help integrate certain types of expression. 

\begin{theorem}[Substitution Rule]\label{theorem:sub}
If $u = g(x)$ is a differentiable function whose range is an interval $I$, and $f$ is continuous on $I$, then
    \begin{equation*}
        \int f\big( g(x) \big)g'(x) dx = \int f(u) du.
    \end{equation*}
\end{theorem}

Like the IBP method, the substitution rule can also be used to generate integrable expressions. We proposed this data generation method to train machine learning models for symbolic integration tasks in  \cite{Barket2024TreeLSTM}. This method has the same requirement as IBP: an existing database of known integrals. This is because the transformation in Theorem \ref{theorem:sub} will only allow us to evaluate the integral if $f$ is integrable. In this case, the database could be comprised of the integrals from the FWD, BWD, and RISCH methods. But that would mean the expressions generated by the SUB method will be somewhat similar to those in the existing datasets since the generator does compositions and derivatives of expressions from the dataset.

\section{Background Material for the New Method}
\label{sec:Background}

The LIOUVILLE generator is motivated in two parts: by Liouville's theorem, and by the adapted version of the Risch algorithm known as the Parallel Risch algorithm \cite{Norman1977_parallel}. Liouville's theorem is the theoretical basis for how these integration algorithms work, and the Parallel Risch algorithm is what the LIOUVILLE generator is modelled upon.

\subsection{Liouville's Theorem}\label{sec:liouville}

Liouville's theorem, and a proof of the theorem, were first given by Risch \cite{Risch1969_original}.

\begin{theorem}[Liouville's theorem:]
    Let $D$ be a differential field with constant field $K$ (whose algebraic closure is $L$). Suppose $f \in D$ and there exists $g$, elementary over $D$, such that $g'=f$. Then, $\exists \, v_0 \in D, c_1,...,c_m \in L, v_1,...,v_m \in D$ such that
    \begin{equation*}
        f=v_{0}' + \sum_{i=1}^{m} c_i\frac{v_{i}'}{v_i} \implies g = v_0 + \sum_{i=1}^{m} c_{i}\log(v_{i}).      
    \end{equation*}
\end{theorem}

Liouville's theorem gives an explicit representation for the \textit{elementary} integral of $f$, if it exists. There have been extensions of Liouville's theorem for various different special functions \cite{Baddoura2006_Liuoville, Kumbhakar2023_liouville, Singer1985_Liouville}. However, there is no general extension of Liouville's theorem that works on all special functions. Not only is this theorem used for the correctness of the original Risch algorithm, but also for the Parallel Risch algorithm discussed next.  It is also the theoretical backbone for our data generation method in Section \ref{sec:New}.

\subsection{Parallel Risch Algorithm}

The original Risch algorithm works by defining a tower $\mathbb{F}(\theta_0,...,\theta_n)$ of differential field extensions over the field  of constants $\mathbb{F}$, where $\theta_i$ can be one of the following:
\begin{enumerate}
    \item \textbf{Logarithmic}: $\theta_i$ = $\log(u)$, $u \in \mathbb{F}(\theta_0,\dots,\theta_{i-1})$.
    \item \textbf{Exponential}: $\theta_i$ = $e^u$, $u \in \mathbb{F}(\theta_0,\dots,\theta_{i-1})$.
    \item \textbf{Algebraic}: $\exists \, p \in \mathbb{F}(\theta_0,\dots,\theta_{i-1})[z]$ such that $p(\theta_i)=0$.
\end{enumerate}
It then recursively builds the integral by finding solutions to sub-problems over simpler domains. The Risch algorithm and proof of its correctness are quite complex, and recursive in nature \cite{Geddes1989_parallelMaple}. 

Because of this complexity, Norman and Moore proposed a new integration algorithm in \cite{Norman1977_parallel}, that works with each extension considered in parallel rather than recursively starting from the top-down\footnote{The paper \cite{Norman1977_parallel} is not currently accessible online. The survey paper \cite{ND79} is probably the earliest reference available online.}.  This has since become known as the \emph{Parallel Risch Algorithm} (or sometimes the Risch-Norman Algorithm). 

The algorithm pseudo-code is reproduced as Algorithm \ref{algo:parallel_risch}.  We present a simplified version of the algorithm and refer the reader to \cite{Davenport1982_parallel, Geddes1989_parallelMaple} for the detailed version. 
Algebraic extensions are not used in this algorithm. Instead, we allow tangent (in addition to logarithmic and exponential)  extensions meaning $\theta_i = \tan(u), u \in \mathbb{F}$. Tangent extensions avoid computation with complex exponentials for
trigonometric functions but note that the original Risch algorithm would do tangent extensions with the complex exponential case.   

\begin{algorithm}[p]
    \caption{The Parallel Risch Algorithm}\label{algo:parallel_risch}
    \textbf{Input:} $f=\frac{p}{q}$ s.t. $p, q \in \mathbb{F}[\theta_0,...,\theta_n] \text{ for } \mathbb{F}$ a constant subfield, $\theta_0=x$, $\theta_i$ is algebraically independent from $\theta_j \text{ when } i \neq j, \text{gcd}(p,q)=1$, and $q$ monic. \\
    \textbf{Output:} The anti-derivative of $f$ or NULL.
    \begin{algorithmic}[1]
    
    \State Perform square-free factorization on $q$ to get $q=\prod_{i=1}^r q_i^{d_i}$ s.t. $q_i$ is square-free.
    \label{step:pr_sqrfree}
    
    \State Let $D=\prod_{i=1}^{r} q_i^{d_{i}^{*}}$ where $d_{i}^{*}=
        \begin{cases}
            d_i,& \text{if $q_i$ is exponential or  $1+\tan^2(\theta)$}, \\
            d_i - 1,& \text{otherwise}.
        \end{cases}
        $
    \label{step:pr_D}
    
    \State Form a hypothesis solution structure based on Liouville's theorem: 
    \[\int \frac{p}{q} d\theta_0 = \frac{U(\theta_0,...,\theta_n)}{D} + \sum c_i\log(v_i).
    \]  
    Here, $U$ is a multivariate polynomial to be determined; and the logarithmic part of the solution, $\sum c_i\log(v_i)$, is formed from the irreducible factors of the $q_i$'s, denoted $v_i$, and coefficients $c_i$ are indeterminates to be assigned later.
    \label{step:pr_structure}
    
    \State Differentiate both sides of Step \ref{step:pr_structure} and clear the denominator. By equating the coefficients of each of the monomials, we obtain a linear system in terms of the unknown coefficients of the monomials in $U$ and the $c_i$'s.
    \label{step:pr_system}
    
    \State Solve the system of equations in Step \ref{step:pr_system} using a linear solver. 
    \label{step:pr_solve}
    
    \If{there is a solution to the system}
        \State \textbf{return} $\frac{U(\theta_0,...,\theta_n)}{D} + \sum c_i\log(v_i)$ after substituting the solution from Step \ref{step:pr_solve}.
    \Else \State \textbf{return} NULL
    \EndIf
    \end{algorithmic}
\end{algorithm}

In Step \ref{step:pr_D}, the exponential and $1+\tan^2(\theta)$ extension cases are treated separately because their degree in $\theta$ does not drop when differentiated. The Parallel Risch algorithm does not always guarantee an output on whether the integral exists or not, unlike the original Risch algorithm. There has been progress in proving certain cases will guarantee a solution, such as \cite{Davenport1986d_parallelCases}, with the work of Bronstein \cite{Bronstein2007_parallel} the most comprehensive to date.  For our use of the Parallel Risch algorithm, we are constructing the solution ourselves so this lack of a guarantee does not cause us a barrier.  The Parallel Risch algorithm motivated the structure of our generator which we present next.

\section{The LIOUVILLE Data Generation Method}
\label{sec:New}

\begin{algorithm}[p]
    \caption{Overview of Liouville Generator}\label{algo:liouville_gen}
    \textbf{Input:} List of field extensions $T=[\theta_0,...,\theta_n]$ where $\theta_0=x$. \\
    An upper bound on the multiplicity of the denominator $r$. \\
    A boolean flag \texttt{normal} that determines whether to return the answer in normalised or partial fraction form. \\
    \textbf{Output:} $F', F$ such that $F'$ is an integrand and $F$ is its integral.
    \begin{algorithmic}[1]
    
    \State Generate polynomials $q_1,...,q_r$ in $\theta_n$ with coefficients in $\theta_0,...,\theta_{n-1}$.
    \label{step:gen_qs}
    
    \State Let $D= \texttt{SquareFreeFactor}(q_1^1 \dots q_{r}^{r}) = Q_1^1 \dots Q_s^s$ where $s \geq r$. \Comment{Denominator}
    \label{step:gen_D}
    
    \State Generate polynomial $N$ in $\theta_n$ with coefficients in $\theta_0,...,\theta_{n-1}$ which has total degree smaller or equal to the total degree of $D$. \Comment{Numerator}
    \label{step:gen_N}
    
    \State Choose $j \leq s$. Generate $a_0,...,a_j \in \{Q_1,...,Q_s\}$ and $c_0,...,c_j \in \mathbb{F}$. Let $A~=~\sum_{i=0}^j c_i\log{(a_i)}$.
    \label{step:gen_as}
    
    \State Choose $k \in \mathbb{N}_0$. Generate $b_0,...,b_k \in \mathbb{F}[\theta_0,...,\theta_n]$ and $d_0,...,d_k \in \mathbb{F}$.  Let $B~=~\sum_{i=0}^k d_i\log{(b_i)}$.
    \label{step:gen_bs}

    \If{\texttt{normal}}
    \State $\hat{F} = \frac{N}{D}$;
    \label{step:gen_F_norm1}
    \State \textbf{return} $\texttt{Normalise}(\hat{F}' + A') + B',  \texttt{Normalise}(\hat{F} + A) + B$
    \label{step:gen_F_norm2}
    \label{step:gen_return_norm}
    \Else
    \State $G = \texttt{PartialFraction}(\frac{N}{D}) + A + B$
    \label{step:gen_F_PF}
    \State \textbf{return} $G', G$
    \EndIf
    
    \end{algorithmic} 
    
\end{algorithm}

\subsection{The Main Idea}
\label{sec:New_algo}
We now present the LIOUVILLE data generation method in Algorithm \ref{algo:liouville_gen}. Note:
\begin{itemize}
    \item \texttt{SquareFreeFactor} refers to the square-free factorization procedure. For example, the Maple function \texttt{sqrfree}: \url{https://www.maplesoft.com/support/help/maple/view.aspx?path=sqrfree}.
    \item \texttt{Normalise} refers to normalising an expression to have a common denominator.  For example, the Maple function \texttt{normal}: \url{https://www.maplesoft.com/support/help/maple/view.aspx?path=normal}
    \item \texttt{PartialFraction} refers to putting an expression in its partial fraction representation. For example, the Maple function \texttt{convert} called with parameter \texttt{parfrac}: \url{https://www.maplesoft.com/support/help/maple/view.aspx?path=convert%2Fparfrac}
\end{itemize}

The design of the data generator has similar properties to the Parallel Risch algorithm (Algorithm \ref{algo:parallel_risch}).  We will walk through the steps to see how they relate. 
\begin{itemize}
    \item We first generate a denominator $D$ and perform square-free factorization to ensure the denominator of our new integral has the same property as ensured in Step \ref{step:pr_sqrfree} and \ref{step:pr_D} of Algorithm \ref{algo:parallel_risch}.
    
    \item We then generate the numerator which is supposed to mimic $U$ in Step \ref{step:pr_structure} of Algorithm \ref{algo:parallel_risch}.
    
    \item Steps \ref{step:gen_as} and \ref{step:gen_bs} of Algorithm \ref{algo:liouville_gen} are distinct in that Step \ref{step:gen_as} chooses arguments of the logarithms directly from the factors of the denominator $\{Q_1,\dots, Q_s\}$, and then Step \ref{step:gen_bs} allows any factors from the field as arguments. The logarithms in Steps \ref{step:gen_as} and \ref{step:gen_bs} are to get denominators in the partial fraction decomposition of the integrand where they have degree 1. Some of the denominators of multiplicity 1 should be the same as the factors of D (Step \ref{step:gen_as}) and some should be different (Step \ref{step:gen_bs}). 
    \\ These two key steps are what separate this generator from the BWD method. The BWD method would have a hard time generating fractions of denominator degree 1 in the partial fraction representation of an expression.  
    
    \item We normalise $\frac{N}{D} + A$ (if requested) in Steps \ref{step:gen_F_norm1} and \ref{step:gen_F_norm2} of the LIOUVILLE generator to control the length of the integrand and integral. See Section \ref{sec:discussion_normalisation} for a thorough discussion on this. 
\end{itemize}

\subsection{Design Choices}

Although we only present Algorithm \ref{algo:liouville_gen} to have three inputs, there are further design choices that depend on the implementation. In these subsequent sections, the goal is to use these choices to create ``\emph{realistic}'' looking expressions rather than just random, complicated integrands. We separate these discussions from Algorithm \ref{algo:liouville_gen} as these are design choices independent of the key idea.  I.e. the rules below do not need to be followed for Algorithm \ref{algo:liouville_gen} to work. We suggest these design choices based on the experience of the third author on what CAS users most commonly require.  

\subsubsection{Choosing $\theta_i$:} Each $\theta_i$ will be of the form $f(u)$ for $f, u \in \mathbb{F}(\theta_0,...,\theta_{i-1})$. To create each $\theta_i$ one can use a random expression generator (such as the one in \cite{Lample2020}). The outer function $f$ can be elementary, algebraic, or special but we wish to avoid deeply nested composition of functions (very rarely should a nesting depth higher than two ever occur). The inner function $u$ can be a random rational function of all the previous indeterminates $\theta_{i-1},...,\theta_0$.

\subsubsection{Generation of $q_i$'s:} In Step \ref{step:gen_qs}, we should be careful to restrict how big the coefficients are for $q_i$. In theory, one can create unnecessarily complicated $q_i$'s by using every single previous extension $\theta_{i-1}$ to $\theta_0$ in the coefficient of $q_i$. However, we still aim for the $q_i$'s to be similar to CAS user inputs. Thus, the coefficient of each $q_i$ should usually have one and very rarely more than one of {$\theta_0,...,\theta_{n-1}$} in the coefficient. For special function extensions, we want these to be much more infrequent than elementary extensions when sampling for the coefficients.

\subsubsection{Creating the Numerator:} Most of the numerators created should be of lower degree in $\theta_n$ than the denominator. If they were greater or equal to the degree in the denominator, we can perform polynomial division and have a polynomial and proper rational part. However, the polynomial part is already similar to the BWD method from Lample \& Charton \cite{Lample2020} as we discussed in  \cite{Barket2023_generation}. Therefore, we mostly want to create proper rational expressions to begin with to diversify our existing dataset.

\section{Discussion of the New Data Generation Method}
\label{sec:discussion}

In this section, we analyse how the LIOUVILLE generator differs from the prior methods introduced in Section \ref{sec:existing}. At first glance, the LIOUVILLE method appears to be a glorified version of the BWD method from \cite{Lample2020} since we are creating an expression and taking its derivative! Further, since the generator also follows from Liouville's theorem, we might also assume it created expressions similar to the RISCH method from \cite{Barket2023_generation}. We will explain below that the LIOUVILLE method does in fact have some key benefits over both BWD and RISCH. 

\subsection{Example}

First, to show the strength of the LIOUVILLE generator, let us revisit Equation (\ref{eq:setup_new}) from earlier and walk through how the generator may produce an answer similar to Equation (\ref{eq:diff_setup_new}). 

\begin{enumerate}
    \item The input to Algorithm \ref{algo:liouville_gen} would be $T=[\theta_0=x, \theta_1=\log(x)], r=2$ and we may suppose the flag is set to \texttt{normal}=\texttt{True}.
    
    \item Through a random process, we generate $q_1=1, q_2=\theta_{1}$ in Step \ref{step:gen_qs}.
    
    \item We perform square-free factorization on $q_{1}q_{2}^2$ in Step~\ref{step:gen_D}. Since $q_2=\log(x)$ is the only actual factor and is linear, it is already in square-free form. Thus $D=\log(x)^2$.
    
    \item Through a random process, we generate the numerator $N=\theta_1 + 1$ in Step~\ref{step:gen_N}. 
    
    \item Choose $j=1$ in Step \ref{step:gen_as} of the algorithm. We must choose $a_0 \in \{\log(x) \}$, i.e. we have only one choice: $a_0=\log(x)$.  We choose $c_0=1$, so $A=\log(\log(x))$.
    
    \item We choose $k=0$ in Step \ref{step:gen_bs} of the algorithm, so $B=0$. We did this to make the example simple but in practice we would usually have $k > 0$.
    
%

    \item The input requested us to normalise and so we calculate  
    \begin{align*}
        \hat{F} &= \frac{N}{D} = \frac{ \log(x)+1}{\log^2(x)} \\
        \hat{F}' &= \left(\frac{1}{x\log(x)^2}\right) 
        - 2\left(\frac{(\log(x)+1)}{x\log(x)^3}\right) \\
        A' &= \frac{1}{x\log(x)}, \qquad 
        B' = 0
    \end{align*}
    and form the integrand
    \[
    \texttt{Normalise}(\hat{F}' + A') + B' = \frac{\log^2(x) - \log(x) - 2}{x\log^3(x)},
    \]
    and integral
    \[
    \texttt{Normalise}(\hat{F} + A) + B 
    = \left( \frac{\log(\log(x))\log(x)^2 + \log(x) + 1}{\log^2(x)} \right).
    \]
\end{enumerate}

If the flag had alternatively been \texttt{normal}=\texttt{False} we would have instead calculated integral
    \[
    G = \texttt{PartialFraction}\left(\frac{N}{D}\right) + A + B
    = \frac{1}{\log(x)} + \frac{1}{\log^2(x)} + \log(\log(x)),
    \]
    and integrand
    \[
    G' = -\frac{1}{x \log(x)^2} - \frac{2}{x\log(x)^3} + \frac{1}{x\log(x)}.
    \]

We have the option of producing the integrant and integral in both their normalised form and their partial fraction form since there is a benefit to having each. We discuss in Section \ref{sec:discussion_normalisation} how the normalised form helps with controlling the length (in terms of the number of tokens) of the integrand. The partial fraction form demonstrates the ability to get degree 1 denominators for one of the fractions, which is something BWD has a very hard time achieving. Depending on the use of the dataset (such as ML), it may be better to actually produce \textit{both} versions of the integrand in a generated dataset.

\subsection{Benefits over FWD and BWD}
\label{sec:discussion_FWDBWD}

\begin{figure}[htbp]
    \centering
    \begin{subfigure}[b]{0.90\textwidth}
        \centering
        \includegraphics[width=\textwidth]{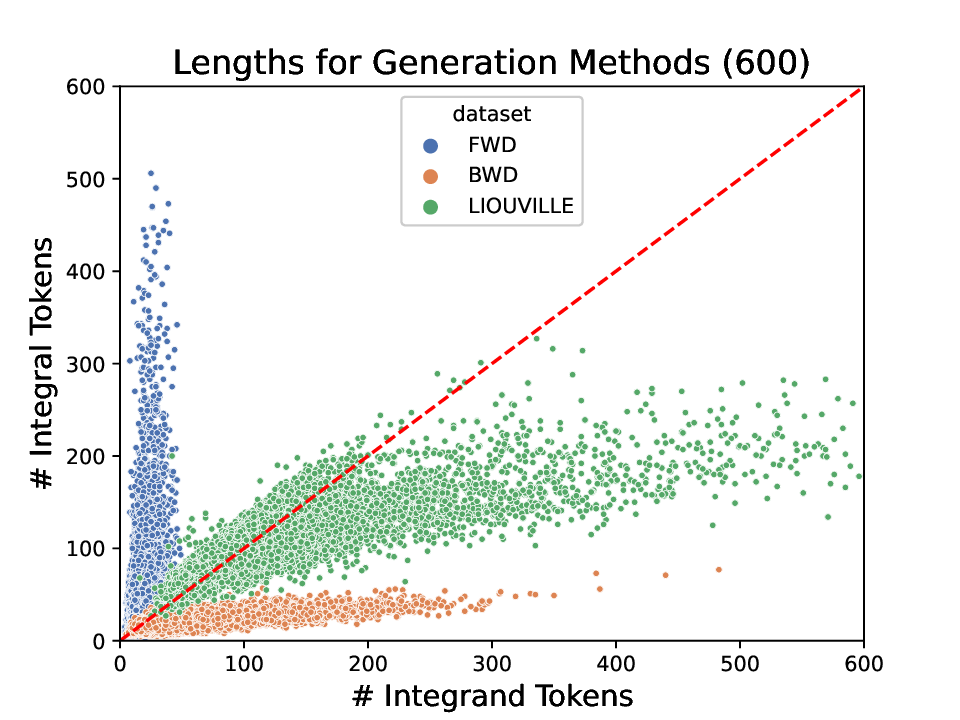}
        \caption{Lengths up to 600 tokens}
        \label{fig:lengths_600}
    \end{subfigure}
    \begin{subfigure}[b]{0.90\textwidth}
        \centering
        \includegraphics[width=\textwidth]{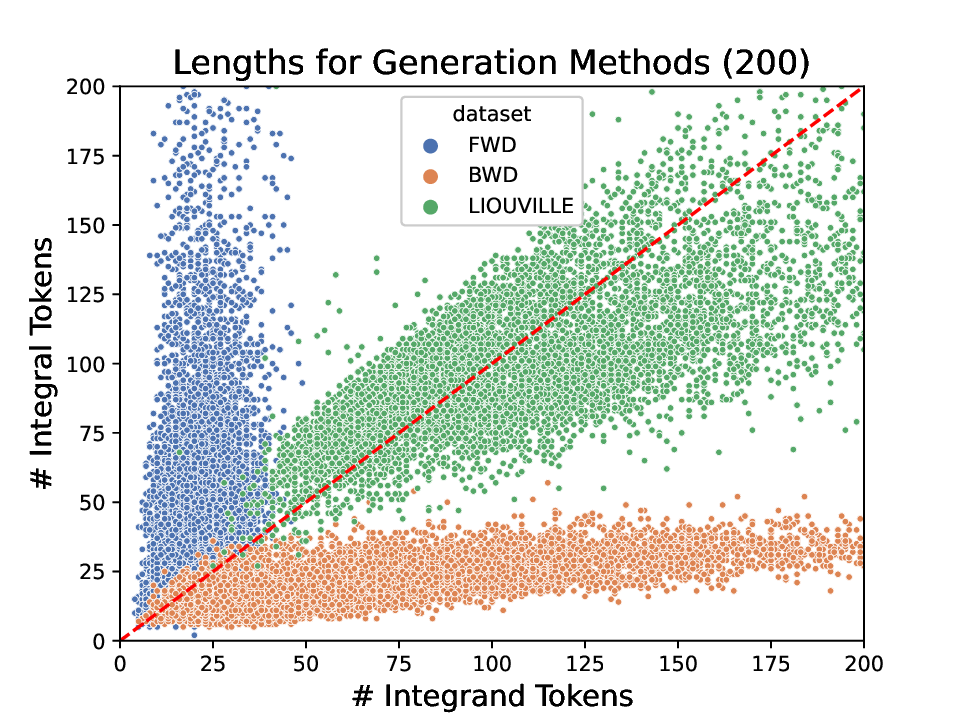}
        \caption{Lengths up to 200 tokens}
        \label{fig:lengths_200}
    \end{subfigure}
    \caption{Comparing the size of the integrands vs. integrals for the FWD, BWD, and LIOUVILLE generators. The closer a point is to the dotted red line $y=x$, the more balanced a method is. Figure \ref{fig:lengths_200} is a zoomed-in portion of Figure \ref{fig:lengths_600}.}
    \label{fig:length_comparison}
\end{figure}

To see how the method combats the length bias shown in the BWD method, we construct a sample set of 10,000 samples from the FWD, BWD, and LIOUVILLE generators and plot the lengths (in prefix notation) of the integrands and integrals in Figure \ref{fig:length_comparison}. This shows that LIOUVILLE does a better job at balancing the lengths between the two compared to the FWD and BWD methods. There is still a bias towards longer integrands than integrals from this method: this is to be expected because the procedure to create (integrand, integral) pairs is a similar, but improved, version of BWD; however the bias is greatly reduced.

In addition, we perform the same analysis as \cite{Piotrowski2019_critique} to see how similar expressions are up to their coefficients. In a sample of 100,000 points, we find that \textit{none} of the generated integrands are similar to each other, a major strength over the FWD, BWD, and IBP methods in \cite{Lample2020}. Again, this is to be expected since there are many randomized parameters within the algorithm: the generation of the extensions, the number of factors, and the inclusion of a variable amount of logarithms. It would be extremely unlikely to generate two similar expressions given the amount of parameters in the algorithm.  

\subsection{Benefits over RISCH}
\label{sec:disccusion_risch}
What if instead we had $q_1=\theta_{1}^3 + x$ and $q_2=1$? Our denominator would then be $D=\log^3(x) + x$ and we would run through the rest of the steps with no issue. However, this is a major problem for the RISCH generation method. Certain computations in the implementation of the RISCH generation method in \cite{Barket2023_generation} would cause our implementation to halt due to computational difficulty. Although both of these generators follow the structure of Liouville's theorem, our new generator has a much easier time generating higher degree denominators in any of the extensions.

What if we included a special function in $T$, our list of extensions? Further, what if we had three or more extensions in $T$? The RISCH generator can handle these cases in theory, but they are very hard to implement in practice. For one, a large tower of extensions would take a long time because you have to recursively handle each extension before generating a solution. Research has also been done on the Risch algorithm for special functions \cite{Raab2013_special} but this has not been generalised to all special functions yet. The LIOUVILLE generator can handle both of these cases with ease since (a) each extension is handled in parallel and (b) we only need to differentiate a special function: there is no system of equations needing to be solved, something our implementation of the RISCH generator relied on.

The RISCH method did address some of the issues in the FWD, BWD, and IBP methods (as mentioned in Section \ref{sec:risch_gen}). Conversely, the BWD method can deal with issues present in the RISCH method (such as large towers of extensions and special functions). However, the LIOUVILLE generator is able to address the issues that these previous four generators have all at once, while also being conceptually simple and easy to implement. 

\subsection{The Effect of Normalisation}
\label{sec:discussion_normalisation}

\begin{figure}[htbp]
    \centering
    \begin{subfigure}[b]{0.575\textwidth}
        \centering
        \includegraphics[width=\textwidth]{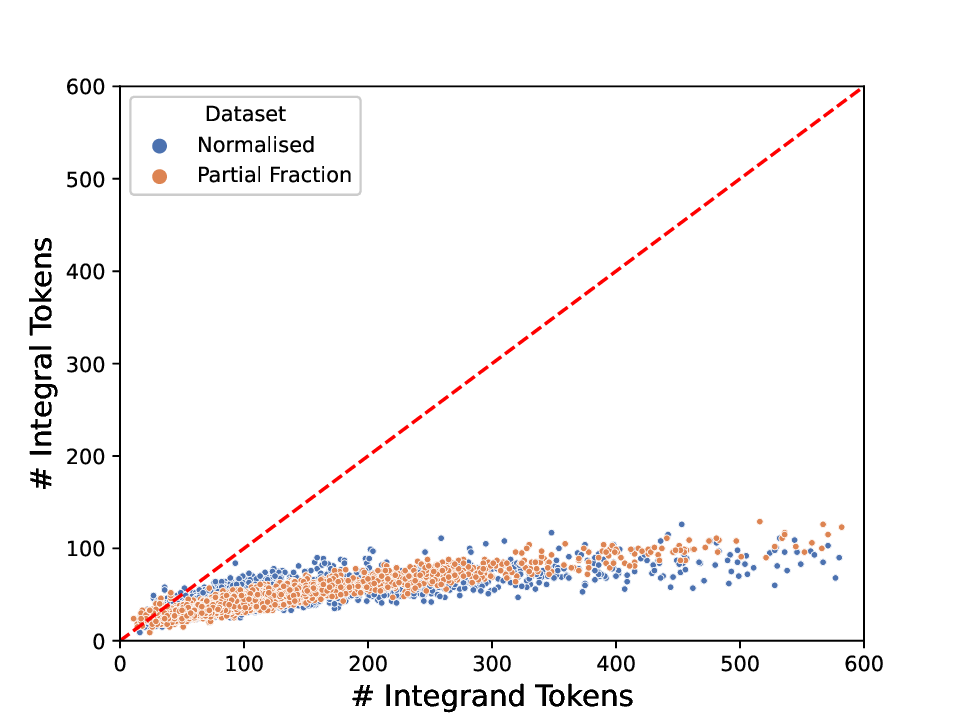}
        \caption{Normalising integrand only}
        \label{fig:normal_integrand}
    \end{subfigure}
    \begin{subfigure}[b]{0.575\textwidth}
        \centering
        \includegraphics[width=\textwidth]{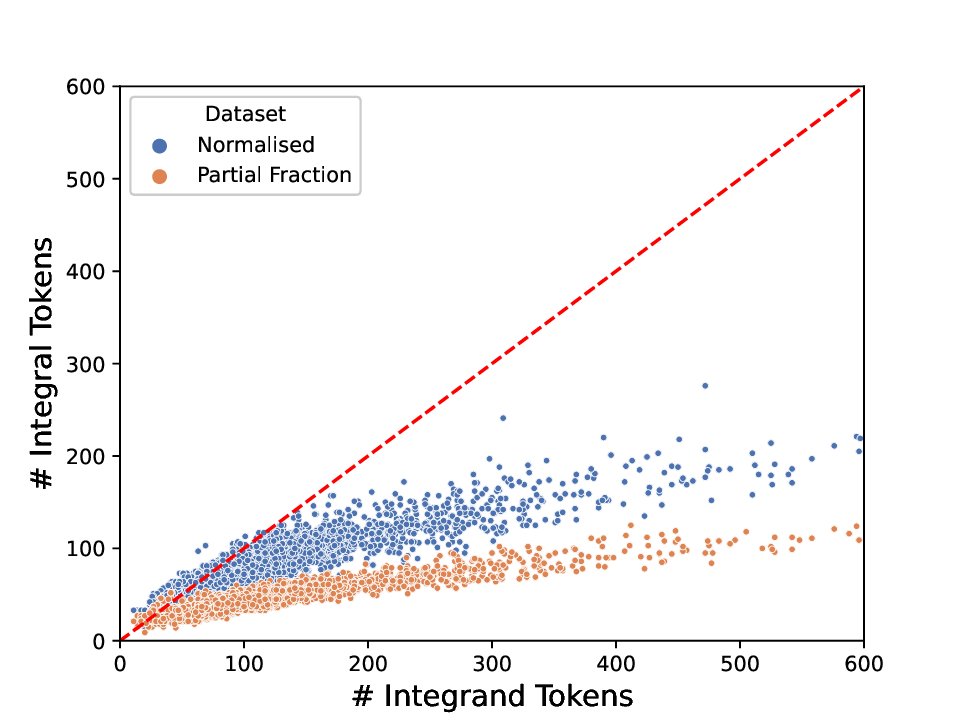}
        \caption{Normalising integral only}
        \label{fig:normal_integral}
    \end{subfigure}
    \begin{subfigure}[b]{0.575\textwidth}
        \centering
        \includegraphics[width=\textwidth]{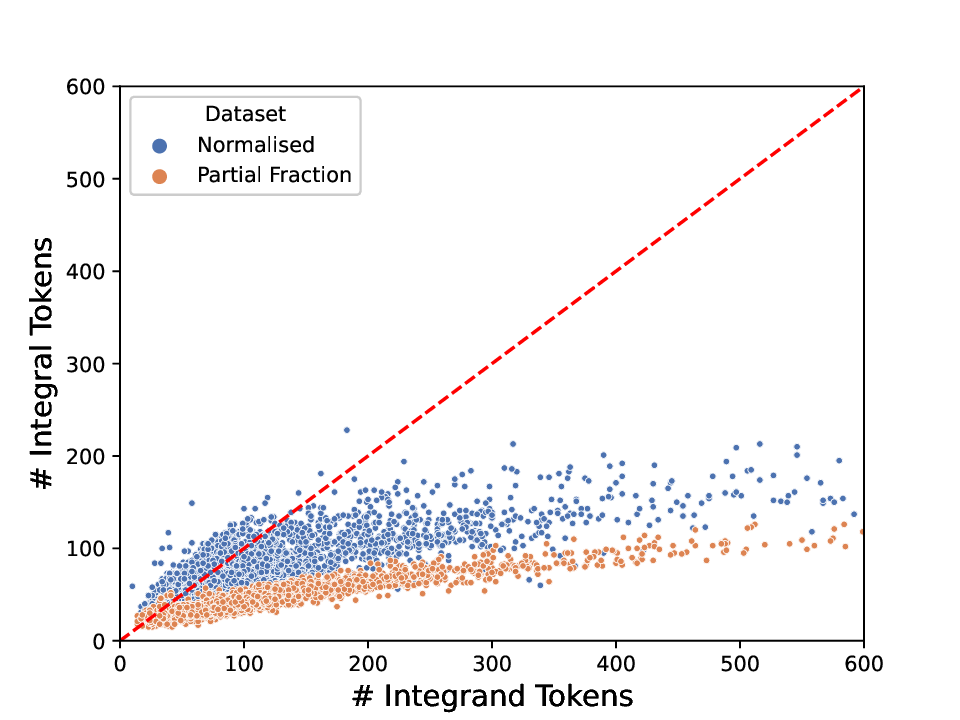}
        \caption{Normalising integrand and integral}
        \label{fig:normal_both}
    \end{subfigure}
    \caption{The effect of normalisation when applied to the integrand, the integral, and both. Normalising both produced more points above the dotted line $y=x$ compared to only normalising one of the integrand or integral.}
    \label{fig:normal_effect}
\end{figure}

In Algorithm \ref{algo:liouville_gen}, there are two different return options: one for a normalised output and one for an expanded (partial fraction) output. The partial fraction view is useful for seeing the degree 1 denominator in $\theta_n$ for one of the fractions. We also state in Section \ref{sec:New_algo} that normalising the output helps to control the length of the integrands and integrals. We empirically show this in Figure \ref{fig:normal_effect} by testing various cases of normalising the integrand and/or the integral. Each case is compared to its partial fraction form (i.e. normalising neither) as a baseline. In these experiments, we set $k=0$ in Step \ref{step:gen_bs} of the LIOUVILLE generator to focus solely on normalisation. 

From Figure \ref{fig:normal_effect}, we can conclude that:
\begin{itemize}
    \item when the integral is normalised, the integral becomes bigger;
    \item when the integrand is normalised, the integrand becomes smaller; and
    \item when both are normalised, integrand and integral sizes are more balanced.  
\end{itemize}
In the case of the integrand, normalising shrinks the size of the integrand because of the way the extra logarithms $A$ are made. When we take the derivative of $A$, $A'$ will have the same factors in its denominator as $D$. Therefore, normalising $\frac{N}{D} + A$ will put everything under a common denominator to limit the growth of the numerator due to the similar factors. 
This also explains why the integral grows bigger after normalisation. Since we did not take the derivative of $A$ yet, $A$ does not have any common factors in its denominator with $D$. Normalising $\frac{N}{D} + A$ makes the numerator grow bigger in size due to $A$ not having any common factors in the denominator. 

Recall that one of the issues with the BWD method is a bias towards big integrands and small integrals (see Figure \ref{fig:length_comparison}). Since normalising has the ability to shrink the integrand and grow the integral, the effect of normalising both of these helps balance the sizes between the two. Observing Figures \ref{fig:normal_integral} and \ref{fig:normal_both}, the benefit of normalising both the integrand and integral starts to become indifferent from normalising just the integral once the integrand size reaches roughly 200 tokens. Algorithm \ref{algo:liouville_gen} will always make the choice of normalising both for simplicity and since there is not a negative effect of doing so.  

\section{Conclusion and Future Work}
\label{sec:Conc}

We presented the LIOUVILLE data generator, a simple method that essentially merges the BWD and RISCH data generation methods and keeps the advantages of both. While simple, the LIOUVILLE generator is able to produce complex and realistic integrands. There is more control over the lengths of the integrand and integral through normalisation, and the generator is extremely unlikely to generate similar expressions as evidenced in Section \ref{sec:discussion_FWDBWD}. Since it is conceptually simpler, it is also easier to implement, allowing us to generate a wider variety of integrals compared to the current implementation of the RISCH method.

Although the Liouville generator has many strengths, it is not perfect. Consider the integral (\ref{eq:easy_int}) introduced earlier. LIOUVILLE, just like BWD, can generate this integrand in theory but it is improbable. Perhaps no one generator will be able to address every edge case of symbolic integration. It is likely a different type of generator or a heuristic added to an existing generator is needed to produce integrands like (\ref{eq:easy_int}) and other such cases. 

The LIOUVILLE generator helped helped in overcoming the size bias in the BWD dataset. However, there does not exist a generator to overcome the bias in the FWD. Examining Figure \ref{fig:length_comparison}, a generator still needs to be invented to create data points between the blue samples (FWD) and the green samples (LIOUVILLE). This way, all data generators combined would be able to handle variable amounts of length in both the integrand and integral.

Nonetheless, the LIOUVILLE generator is a promising method to help produce a rich variety of integrable expressions. The data produced by this generator will help in benchmarking the integration function in a CAS as well as providing more data in ML tasks relating to symbolic integration.

\subsection*{Code Availability Statement} 

A Maple implementation of Algorithm \ref{algo:liouville_gen}, all data used in the paper, and all code to generate the figures are available on Zenodo: \url{https://zenodo.org/doi/10.5281/zenodo.11664457}. We also provide the Github link for the wider project: \url{https://github.com/rbarket/Liouville/tree/main}. 

The Zenodo link is all the current code used for this paper which will remain fixed. At the the GitHub link we will continue to grow the dataset size, make small updates to the algorithm, and continue to add more elementary and special functions to the generator.

\subsection*{Acknowledgements} 

The authors would like to thank James H. Davenport for helpful discussion on the Parallel Risch algorithm.  
Matthew England is supported by EPSRC Project EP/T015748/1, \emph{Pushing Back the Doubly-Exponential Wall of Cylindrical Algebraic Decomposition} (the DEWCAD Project). Rashid Barket is supported by a scholarship provided by Maplesoft and Coventry University.


%
%
\bibliographystyle{splncs04}
\bibliography{ref.bib}
\end{document}